\documentclass{aa} \usepackage{graphicx,epsfig}

\newcommand{\ten}[2]{\mbox{$#1\!\times\!10^{#2}$}}

\def\chem#1#2{\mbox{$\rm{}^{#2}\kern-0.6pt#1$}}
\def\reac#1#2#3#4#5#6{\mbox{$\rm\,{}^{#2}\kern-0.6pt{#1}\,({#3}\,,{#4})\,{}^{#6}\kern-0.6pt{#5}\,$}}
\def\betap#1#2#3#4{$\rm\,{}^{#2}\kern-0.6pt{#1}\,(\beta^+)\,{}^{#4}\kern-0.6pt{#3}\,$}
\def\betam#1#2#3#4{$\rm\,{}^{#2}\kern-0.6pt{#1}\,(\beta^-)\,{}^{#4}\kern-0.6pt{#3}\,$}
\def\reacbp#1#2#3#4#5#6#7#8{$\rm\,{}^{#2}\kern-0.6pt{#1}\,({#3}\,,{#4})\,{}^{#6}\kern-0.6pt{#5}\,(\beta^+)\,{}^{#8}\kern-0.6pt{#7}\,$}
\def\reacbm#1#2#3#4#5#6#7#8{$\rm\,{}^{#2}\kern-0.6pt{#1}\,({#3}\,,{#4})\,{}^{#6}\kern-0.6pt{#5}\,(\beta^-)\,{}^{#8}\kern-0.6pt{#7}\,$}

\def\gsimeq{\,\,\raise0.14em\hbox{$>$}\kern-0.76em\lower0.28em\hbox
{$\sim$}\,\,}
\def\lsimeq{\,\,\raise0.14em\hbox{$<$}\kern-0.76em\lower0.28em\hbox
{$\sim$}\,\,}

\begin{document}

\thesaurus{06(02.14.1)}

\title{Non-explosive hydrogen and helium burnings: Abundance
predictions from the NACRE reaction rate compilation\thanks{An
electronic version of this paper, with colour figures, is available at
{\it http://astro.ulb.ac.be}}}

\author{ M. Arnould \and S. Goriely $^{\star\star}$ \and A. Jorissen
\thanks{Research Associate, F.N.R.S. (Belgium) } }

\institute{Institut d'Astronomie et d'Astrophysique, Universit\'e
Libre de Bruxelles, CP 226, Boulevard du Triomphe, B-1050 Bruxelles,
Belgium}

\offprints{S.~Goriely} \mail{sgoriely@astro.ulb.ac.be}

\date{19 January 1999; 19 April 1999}

\authorrunning{M. Arnould et al.}  \titlerunning{Abundances from
non-explosive H and He burnings}

\maketitle

\begin{abstract}
The abundances of the isotopes of the elements from C to Al produced
by the non-explosive CNO, NeNa and MgAl modes of hydrogen burning, as
well as by helium burning, are calculated with the thermonuclear rates
recommended by the European compilation of reaction rates for
astrophysics (NACRE). The impact of nuclear physics uncertainties on
the derived abundances is discussed in the framework of a simple
parametric astrophysical model. These calculations have the virtue of
being a guide in the selection of the nuclear uncertainties that have
to be duly analyzed in detailed model stars, particularly in order to
perform meaningful confrontations between abundance observations and
predictions. They are also hoped to help nuclear astrophysicists
pinpointing the rate uncertainties that have to be reduced most
urgently.
 
\end{abstract}

\section{Introduction}

The evolution of a star is made of a succession of ``controlled"
thermonuclear burning stages interspersed with phases of gravitational
contraction. The latter stages are responsible for a temperature
increase, the former ones producing nuclear energy and composition
changes.

As is well known, hydrogen and helium burning in the central regions
or in peripheral layers of a star are key nuclear episodes, and leave
clear observables, especially in the Hertzsprung-Russell diagram, or
in the stellar surface composition.  These photospheric abundance
signatures may result from so-called ``dredge-up'' phases, which are
expected to transport the H- or He-burning ashes from the deep
production zones to the more external layers. This type of surface
contamination is encountered especially in low- and intermediate-mass
stars on their first or asymptotic branches, where two to three
dredge-up episodes have been identified by stellar evolution
calculations. Nuclear burning ashes may also find their way to the
surface of non-exploding stars by rotationally-induced mixing, which
has been started to be investigated in some detail (Heger 1998), or by
steady stellar winds, which have their most spectacular effects in
massive stars of the Wolf-Rayet type.
 
The confrontation between the wealth of observed elemental or isotopic
compositions and calculated abundances can provide essential clues on
the stellar structure from the main sequence to the red giant phase,
and much has indeed been written on this subject. Of course, the
information one can extract from such a confrontation is most
astrophysically useful if the discussion is freed from nuclear physics
uncertainties to the largest possible extent.

Thanks to the impressive skill and dedication of some nuclear
physicists, remarkable progress has been made over the years in our
knowledge of reaction rates at energies which are as close as possible
to those of astrophysical relevance (e.g. Rolfs \& Rodney 1988).
Despite these efforts, important uncertainties remain. This relates
directly to the enormous problems the experiments have to face in this
field, especially because the energies of astrophysical interest for
charged-particle-induced reactions are much lower than the Coulomb
barrier energies. As a consequence, the corresponding cross sections
can dive into the nanobarn to picobarn abyss. In general, it has not
been possible yet to measure directly such small cross
sections. Theoreticians are thus requested to supply reliable
extrapolations from the lowest energies attained experimentally to
those of most direct astrophysical relevance.

Recently, a new major challenge has been taken up by a consortium of
European laboratories with the build-up of well documented and
evaluated sets of experimental data or theoretical predictions for a
large number of astrophysically interesting nuclear reactions (Angulo
et al. 1999). This compilation of reaction rates, referred to as NACRE
(Nuclear Astrophysics Compilation of REaction rates; see Sect.~2 for
some details), comprises in particular the rates for all the
charged-particle-induced nuclear reactions involved in the ``cold''
pp-, CNO, NeNa and MgAl chains, the first two burning modes being
essential energy producers, all four being important nucleosynthesis
agents. It also includes the main reactions involved in non-explosive
helium burning.
 
 The aim of this paper is to calculate with the help of the NACRE data
 the abundances of the different isotopes of the elements from C to Al
 involved in the non-explosive H (Sects.~3 - 5) and He (Sect.~6)
 burnings, special emphasis being put on the impact of the reported
 remaining rate uncertainties on the derived abundances. The yields
 from the considered burning modes are calculated by combining in all
 possible ways the lower and upper limits of all the relevant reaction
 rates. One ``reference'' abundance calculation is also performed with
 all the recommended NACRE rates. Note that the pp-chains are not
 considered here. A solar neutrino analysis based on preliminary NACRE
 data for the pp reactions can be found in Castellani et al. (1997).
 
Our extensive abundance uncertainty analysis is performed in the
framework of a parametric model assuming that H burning takes place at
a constant density $\rho = 100$~g~cm$^{-3}$ and at constant
temperatures between $T_6=10$ and 80 ($T_n$ is the temperature in
units of $10^n~{\rm K}$). The corresponding typical values adopted for
He burning are $\rho = 10^4$~g~cm$^{-3}$ and $T_8=1.5$ and 3.5.  These
ranges encompass typical burning conditions in a large variety of
realistic stellar models. For the study of H-burning, initial
abundances are assumed to be solar (Anders \& Grevesse 1989). For
He-burning, we adopt the abundances resulting from H burning at
$T_6=60$ and $\rho = 100$~g~cm$^{-3}$ calculated with the use of the
NACRE recommended rates. The H- and He-burning nucleosynthesis is
followed until the H and He mass fractions drop to $10^{-5}$.

In spite of its highly simplistic aspect, this analysis provides
results that are of reasonable qualitative value, as testified by
their confrontation with detailed stellar model predictions. Most
significant, these parametric calculations have the virtue of
identifying the rate uncertainties whose impact may be of significance
on abundance predictions at temperatures of stellar relevance. They
thus serve as a guide in the selection of the nuclear uncertainties
that have to be duly analyzed in detailed model stars, particularly in
order to perform meaningful confrontations between abundance
observations and predictions. They are also hoped to help nuclear
astrophysicists pinpointing the rate uncertainties that have to be
reduced most urgently.

\section{The NACRE compilation in a nutshell}

A detailed information about the procedure adopted to evaluate each of
the NACRE reaction rates and about the derived values can be found in
Angulo et al.  (1999), or in electronic form at {\it
http://astro.ulb.ac.be}, which also offers the possibility of
generating interactively tables of reaction rates for networks and
temperature grids selected by the user\footnote{This electronic
address also provides many other nuclear data of nuclear astrophysics
interest}. It is clearly impossible to go here into the details of the
NACRE procedure. Let us just emphasize some of its specificities:

\noindent (1)  For
each reaction, the non-resonant and broad-reso\-nan\-ce contributions
to its rate are evaluated numerically in order to avoid the
approximations which are classically made (see Fowler et al. 1975 for
details) in order to allow analytical rate evaluations;

\noindent (2) Narrow or
subthreshold resonances are in general approximated by Breit-Wigner
shapes, and their contributions to the reaction rates are approximated
in the usual analytical way (e.g. Fowler et al. 1975). However, in
some cases, the resonance data are abundant enough to allow a
numerical calculation avoiding these approximations;

\noindent (3) For each reaction, NACRE provides a
recommended ``adopted'' rate, along with realistic lower and upper
limits.  The adopted values of, and the limits on the resonance
contributions are derived from weighted averages duly taking into
account the uncertainties on individual measurements, as well as the
different measurements that are sometimes available for a given
resonance [see Eq.~(15) of Angulo et al. 1999]. For non-resonant
contributions, $\chi^2$-fits to available data provide the recommended
values along with the lower and upper limits, as the experimental
uncertainties on one set of data and the differences between various
sets, if available, are taken into account in the
$\chi^2$-procedure. It is worth stressing at this point that enough
information is provided by NACRE for helping the user to tailor his
own preferred rates if he wants.

The procedure just sketched in (1) - (3) is the selected standard
methodology, and has the advantage of being easily reproducible and of
avoiding any subjective renormalization of different experimental data
sets.Quite clearly, however, the large variety of different situations
makes unavoidable some slight modifications of the standard procedure
in some cases.  These specific adjustments are clearly identified and
discussed in Angulo et al. (1999);

\begin{figure}
   \resizebox{\hsize}{!}{\includegraphics[scale=0.2]{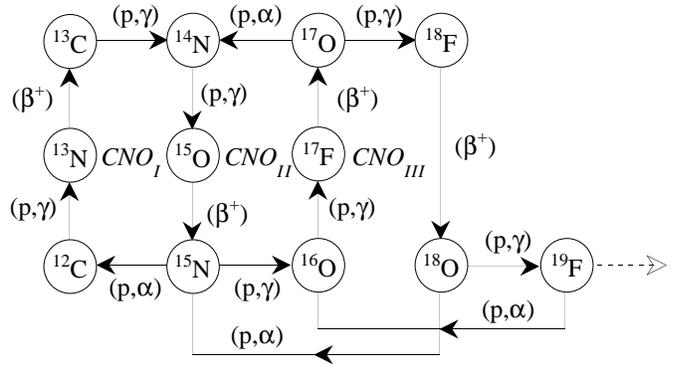}}
  \caption{Reactions of the CNO cycles.     
         The dashed line represents the possible leakage out of the
         cycles }
   \label{fig01}
\end{figure}

\begin{figure*}
   \hspace*{3cm}
   \resizebox{13cm}{!}{\includegraphics[angle=-90]{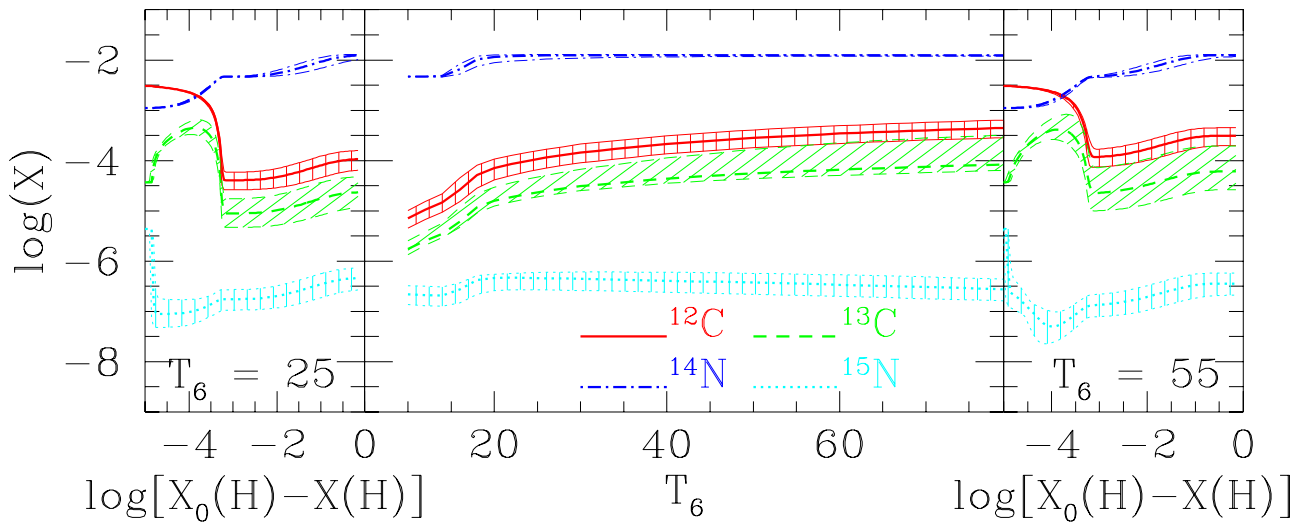}} \hfill
   \vspace{-5cm}

   \caption{ 
          {\it Left and right panels:} Time variations of the mass
          fractions of the stable C and N isotopes versus the amount
          of hydrogen burned at constant density $\rho=100\;{\rm
          g/cm^3}$ and constant temperatures $T_6=25$ and 55. The H
          mass fraction is noted $X$(H), the subscript 0 corresponding
          to its initial value; {\it Middle panel:} Mass fractions of
          the same nuclides at H exhaustion [X(H)=$10^{-5}$] as a
          function of $T_6$. The shaded areas delineate the
          uncertainties resulting from the reaction rates }
   \label{Fig:CNOyields}
\end{figure*}

\noindent (4) A theoretical (Hauser-Feshbach) evaluation of the contribution to each rate of
the thermally populated excited states of the target is also
provided. It has to be noted that the widely used compilation of
Caughlan \& Fowler (1988, hereafter referred to as CF88) provides
uncertainties for some rates only, while the contribution of excited
target states is derived in most cases from a rough (referred to as
``equal strength'') approximation;

\noindent (5) It has to be emphasized that the major goal of the NACRE compilation is to
provide numerical reaction rates in tabular form (see {\it
http://astro.ulb.ac.be}). This philosophy differs markedly from the
one promoted by the previous widely used compilations (CF88, and
references there\-in), and is expected to lead to more accurate rate
data. However, for completeness, NACRE also provides analytical
approximations (Angulo et al. 1999) that differ in several respects
from the classically used expressions (CF88, and references therein).

\section{The CNO Cycles}

\begin{figure*}
   \hspace*{3cm}
   \resizebox{13cm}{!}{\includegraphics[angle=-90]{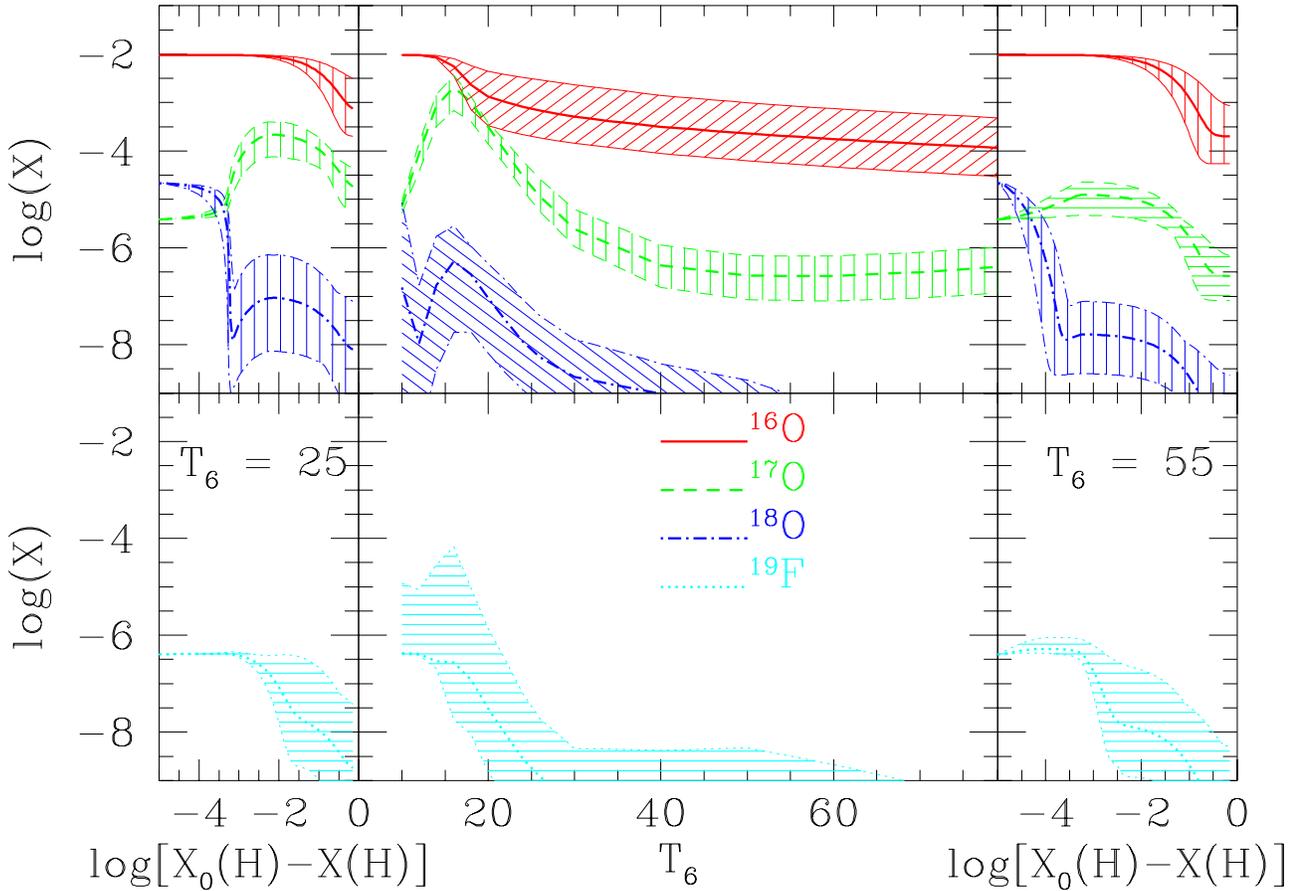}} \hfill
   \caption{Same as Fig.~\ref{Fig:CNOyields}, but for the O and F nuclides}
   \label{Fig:OFyields}

\end{figure*}

The reactions of the CNO cycles are displayed in Fig.~1. As is well
known, their net result is the production of \chem{He}{4} from H, and
the transformation of the C, N and O isotopes mostly into \chem{N}{14}
as a result of the relative slowness of \reac{N}{14}{p}{\gamma}{O}{15}
with respect to the other involved reactions. This \chem{N}{14}
build-up is clearly seen in Fig.~2.

As shown in Fig.~1, three nuclides are important branching points for
the CNO cycles. The first one is \chem{N}{15}.  At $T_6=25$,
\reac{N}{15}{p}{\alpha}{C}{12} is 1000 times faster than
\reac{N}{15}{p}{\gamma}{O}{16}, and the CN cycle reaches equilibrium
already before $10^{-3}$ of the initial protons have been burned. The
second branching is at \chem{O}{17}. The competing reactions
\reac{O}{17}{p}{\alpha}{N}{14} and \reac{O}{17}{p}{\gamma}{F}{18}
determine the relative importance of cycle II over cycle III
(Fig.~\ref{fig01}). The uncertainties on these rates have been
strongly reduced in the last years. The rate of
\reac{O}{17}{p}{\alpha}{N}{14} recommended by NACRE is larger than the
CF88 one by factors of 13 and 90 at $T_6=20$ and 80, respectively.
Smaller deviations, though reaching a factor of 9 at $T_6=50$, are
found for the \reac{O}{17}{p}{\gamma}{F}{18} rate.
  
\begin{figure*}
   \hspace*{3cm}
   \resizebox{13cm}{!}{\includegraphics[scale=0.30,angle=-90]{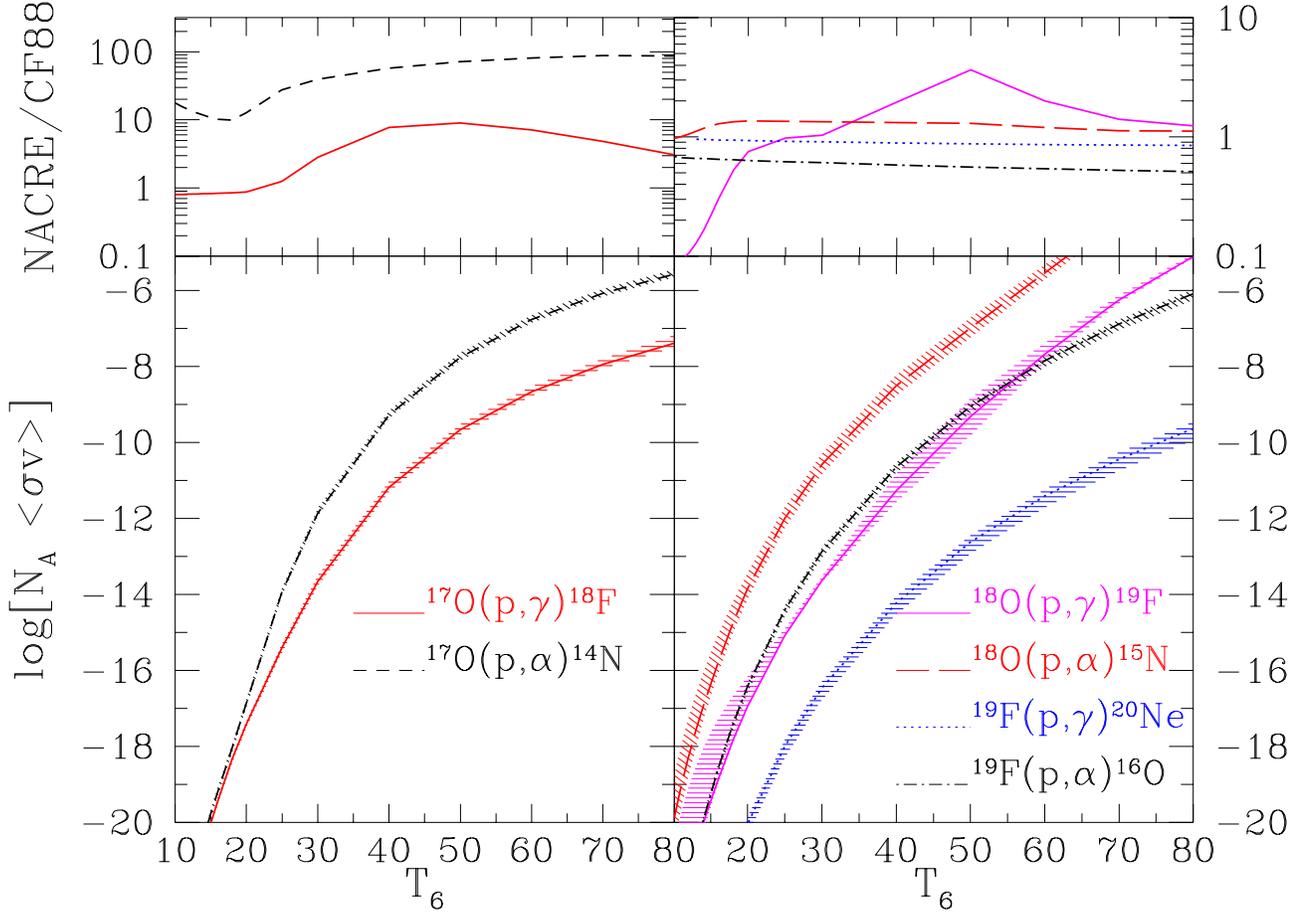}}
   \caption{
         Bottom panels: Temperature dependence of Maxwellian-averaged
         reaction rates (expressed in cm$^3$ mol$^{-1}$ s$^{-1}$) from
         NACRE for proton capture by \chem{O}{17} (left panel) and by
         \chem{O}{18} and \chem{F}{19} (right panel).  The rate
         uncertainties given by NACRE are represented by the shaded
         area. Top panels: Ratio between the NACRE and CF88
         Maxwellian-averaged reaction rates}
   \label{Fig:O17pag}
\end{figure*}

The oxygen isotopic composition is shown in Fig.~3. As it is well
known, it depends drastically on the burning temperature.  In
particular, \chem{O}{17} is produced at $T_6 \lsimeq 25$, but is
destroyed at higher temperatures. This has the important consequence
that the amount of \chem{O}{17} emerging from the CNO cycles and
eventually transported to the stellar surface is a steep function of
the stellar mass. This conclusion could get some support from the
observation of a large spread in the oxygen isotopic ratios at the
surface of red giant stars of somewhat different masses (Dearborn
1992, and references therein). Fig.~3 also demonstrates that the
oxygen isotopic composition cannot be fully reliably predicted yet at
a given temperature as a result of the cumulative uncertainties
associated with the different production and destruction rates.
 
Finally, the leakage from cycle III is determined by the ratio of the
\reac{O}{18}{p}{\gamma}{F}{19} and \reac{O}{18}{p}{\alpha}{N}{15}
rates (Fig.~\ref{fig01}).  At the temperatures of relevance,
\reac{O}{18}{p}{\gamma}{F}{19} is roughly 1000 times slower than
\reac{O}{18}{p}{\alpha}{N}{15}, in relatively good agreement with CF88
(Fig.~\ref{Fig:O17pag}), undermining the path leading to the
production of \chem{F}{19}. However, at low temperatures, large
uncertainties still affect the \reac{O}{18}{p}{\gamma}{F}{19} rate. In
fact, its upper bound could be comparable to the
\reac{O}{18}{p}{\alpha}{N}{15} rate, and at the same time larger than
the \reac{F}{19}{p}{\alpha}{O}{16} rate at $T_6 \lsimeq 20$. As a
result, some \chem{F}{19} might be produced, in contradiction with the
conclusion drawn from the adoption of the CF88 rates. Fig.~3 indeed
confirms that fluorine could be overproduced (with respect to solar)
by up to a factor of 100 at H exhaustion when $T_6 \approx 15$.
However, Fig. 3 also reveals that the maximum \chem{F}{19} yields that
can be attained remain very poorly predictable as a result of the rate
uncertainties.  In fact, some hint of a non-negligible production of
fluorine by the CNO cycles might come from the observation of fluorine
abundances slightly larger than solar at the surface of red giant
stars considered to be in their post-first dredge-up phase (Jorissen
et al. 1992; Mowlavi et al. 1996).

As far as \reac{O}{18}{p}{\alpha}{N}{15} is concerned, let us also
mention that Huss et al. (1997) have speculated that its rate could be
about 1000 times larger than the one adopted by CF88 and NACRE at
temperatures of about $15\times 10^6$ K. This proposal has been made
in order to explain the N isotopic composition measured in some
presolar grains. It is clearly fully incompatible with the NACRE
analysis.

Finally, let us note that \reac{F}{19}{p}{\alpha}{O}{16} is always
much faster than \reac{F}{19}{p}{\gamma}{Ne}{20}. Any important
leakage out of the CNO cycles to \chem{Ne}{20} is thus prevented, this
conclusion being independent of the remaining rate uncertainties.

\section{The NeNa Chain} 

\begin{figure}[h]  
\resizebox{\hsize}{!}{\includegraphics[scale=0.34]{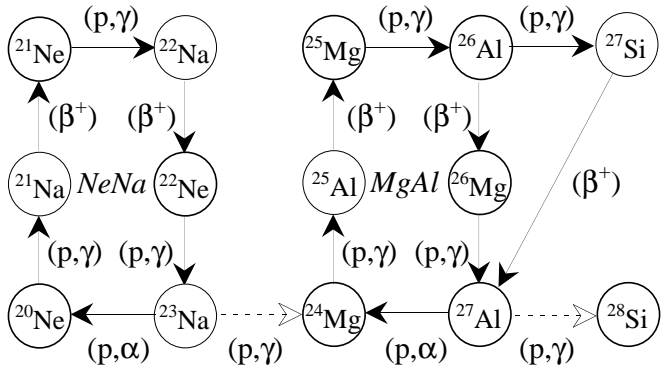}}
   \caption[]{
\label{Fig:NeNa}
Same as Fig.~1, but for the NeNa and MgAl chains}
\end{figure}

\begin{figure*}
   \hspace*{3cm}
   \resizebox{13cm}{!}{\includegraphics[scale=0.30,angle=-90]{H1352.f6}}
   \caption{\label{Fig:Ne21Ne22pg}
          Same as Fig.~\ref{Fig:O17pag}, but for
          \reac{Ne}{20}{p}{\gamma}{Na}{21},
          \reac{Ne}{21}{p}{\gamma}{Na}{22},
          \reac{Ne}{22}{p}{\gamma}{Na}{23} (left panel) and
          \reac{Na}{23}{p}{\gamma}{Mg}{24},
          \reac{Na}{23}{p}{\alpha}{Ne}{20} (right panel) }
\end{figure*}

The NeNa chain is illustrated in Fig.~\ref{Fig:NeNa}, while
Fig.~\ref{Fig:Ne21Ne22pg} displays some relevant NACRE reaction rates,
and their, sometimes quite large, uncertainties. These affect in
particular the proton captures by \chem{Ne}{21}, \chem{Ne}{22} and
\chem{Na}{23}. In contrast, the \reac{Ne}{20}{p}{\gamma}{Na}{21} rate
may be considered as relatively well determined.  Some of these rates
may also deviate strongly from the CF88 proposed values.

\begin{figure*}
   \hspace*{3cm}
   \resizebox{13cm}{!}{\includegraphics[angle=-90]{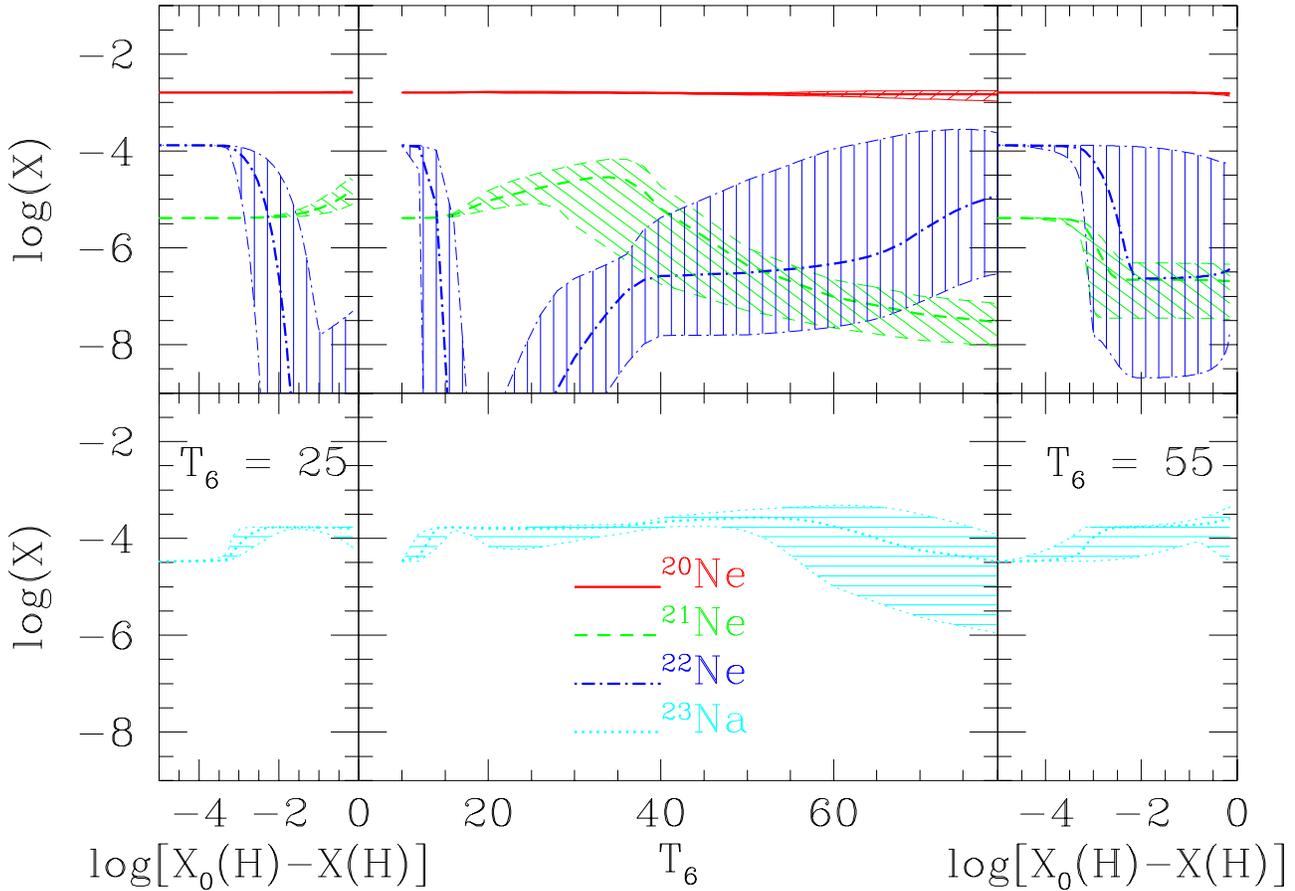}} \hfill
   \caption[~NeNa yields]
        {Same as Fig.~\ref{Fig:CNOyields}, but for the nuclides
        involved in the NeNa chain}
   \label{Fig:NeNayields}
\end{figure*}

The NACRE rates are used to compute the abundances shown in
Fig.~\ref{Fig:NeNayields}. A slight alteration of the initial
\chem{Ne}{20} abundance is visible only for $T_6 \gsimeq$ 50. However,
an unnoticeable \chem{Ne}{20} destruction is sufficient to lead to a
significant increase of the abundance of the rare \chem{Ne}{21}
isotope through \reacbp{Ne}{20}{p}{\gamma}{Na}{21}{Ne}{21} at $T_6
\lsimeq 40$.  At higher temperatures,
\reacbp{Ne}{21}{p}{\gamma}{Na}{22}{Ne}{22} destroys \chem{Ne}{21}.  As
a result, the \chem{Ne}{21} abundance at H exhaustion is maximum when
H burns in the approximate $30 \lsimeq T_6 \lsimeq 35$ range. This
conclusion may, however, be altered if the upper limit of the
\reac{Ne}{21}{p}{\gamma}{Na}{22} rate is adopted instead.
 
The \chem{Na}{23} yield has raised much interest recently, following
the discovery at the surface of globular cluster red giant stars of
moderate sodium overabundances which correlate or anti-correlate with
the amount of other elements (like C, N, O, Mg and Al) also involved
in cold H burning (Denissenkov et al. 1998; Kraft et al. 1998, and
references therein). This situation may be the signature of the
dredge-up to the stellar surface of the ashes of the NeNa chain. The
\chem{Na}{23} production results from
\reac{Ne}{22}{p}{\gamma}{Na}{23}, while
\reac{Na}{23}{p}{\gamma}{Mg}{24} and \reac{Na}{23}{p}{\alpha}{Ne}{20}
are responsible for its destruction, which can be substantial at $T_6
\gsimeq 60$. Unfortunately, our knowledge of these three reaction
rates remains very poor, with uncertainties that can amount to factors
of about 100 to $10^4$ in certain temperature ranges (see
Fig.~\ref{Fig:Ne21Ne22pg}). As indicated in Fig.~\ref{Fig:NeNayields},
this situation prevents an accurate prediction of the \chem{Na}{23}
yields when $T_6 \gsimeq 50$. More precisely, the spread in the
\chem{Na}{23} abundance at H exhaustion reaches a factor of 100 at
these temperatures.

The possible cycling character of the NeNa chain is determined by the
ratio of the rates of \reac{Na}{23}{p}{\alpha}{Ne}{20} and of
\reac{Na}{23}{p}{\gamma}{Mg}{24}. Fig.~\ref{Fig:Ne21Ne22pg} indicates
that the former reaction is predicted to be faster than the latter one
at $T_6 \lsimeq 50$ only. In this case, the NeNa chain is indeed a
cycle. However, at higher temperatures, an important leakage to the
MgAl chain can be expected, unless future experiments confirm the
lower bound of the uncertain \reac{Na}{23}{p}{\gamma}{Mg}{24} rate.

\section{The MgAl Chain}

The MgAl chain is depicted in Fig.~5. It involves in particular
\chem{Al}{26}. Its long-lived ($t_{1/2} =$ \ten{7.05}{5} y)
\chem{Al^g}{26} ground state and its short-lived ($t_{1/2} = 6.35$ s)
\chem{Al^m}{26} isomeric state are out of thermal equilibrium at the
temperatures of relevance for the non-explosive burning of hydrogen
(Coc \& Porquet 1998). They have thus to be considered as separate
species in abundance calculations.

The status of our present knowledge of some important reactions of the
MgAl chain is depicted in Fig.~\ref{Fig:Alg26pg}, while the yield
predictions for the species involved in this chain are presented in
Fig.~\ref{Fig:MgAlyields}. Let us first discuss the situation
resulting from the use of the NACRE adopted rates. The most abundant
nuclide is \chem{Mg}{24}, the concentration of which remains
unaffected, at least for $T_6 \lsimeq 60$. In contrast, \chem{Mg}{25}
is significantly transformed by proton captures into \chem{Al^g}{26}
at $T_6 \gsimeq 30$.  At $T_6 \gsimeq 50$, the leakage from the NeNa
cycle starts affecting the MgAl nucleosynthesis through a slight
increase of the \chem{Mg}{24} abundance, followed by a modest
enhancement of the \chem{Mg}{25}, \chem{Al^g}{26} and \chem{Al}{27}
concentrations (Fig.~\ref{Fig:MgAlyields}). At temperatures $T_6
\gsimeq 70$, the \chem{Mg}{24} accumulation starts turning into a
depletion by proton captures, which contributes to a further increase
in the \chem{Mg}{25}, \chem{Al^g}{26} and \chem{Al}{27}
abundances. This build-up cannot be significantly hampered by the
destruction of these species by proton captures, as a result of their
relative slowness. Among these reactions,
\reac{Al}{27}{p}{\alpha}{Mg}{24} and \reac{Al}{27}{p}{\gamma}{Si}{28}
are of special interest, as the ratio of their rates determines in
particular the leakage out of the MgAl chain. The adopted NACRE rate
of the former reaction is 20 to 100 times slower than the CF88 one in
the considered temperature range, and turns out to be slower than the
(p,$\gamma$) channel for $T_6 \gsimeq 60$ (Fig.~\ref{Fig:Alg26pg}), so
that no cycling back is possible in these conditions.

\begin{figure*}
   \hspace*{3cm}
   \resizebox{13cm}{!}{\includegraphics[scale=0.30,angle=-90]{H1352.f8}}
   \caption{  \label{Fig:Alg26pg}
         Same as Fig.~\ref{Fig:O17pag}, but for
         \reac{Mg}{26}{p}{\gamma}{Al}{27},
         \reac{Al^g}{26}{p}{\gamma}{Al}{27} (left panel) and
         \reac{Al}{27}{p}{\gamma}{Si}{28},
         \reac{Al}{27}{p}{\alpha}{Mg}{24} (right panel) }
\end{figure*}

\begin{figure*}
   \hspace*{3cm}
   \resizebox{13cm}{!}{\includegraphics[angle=-90]{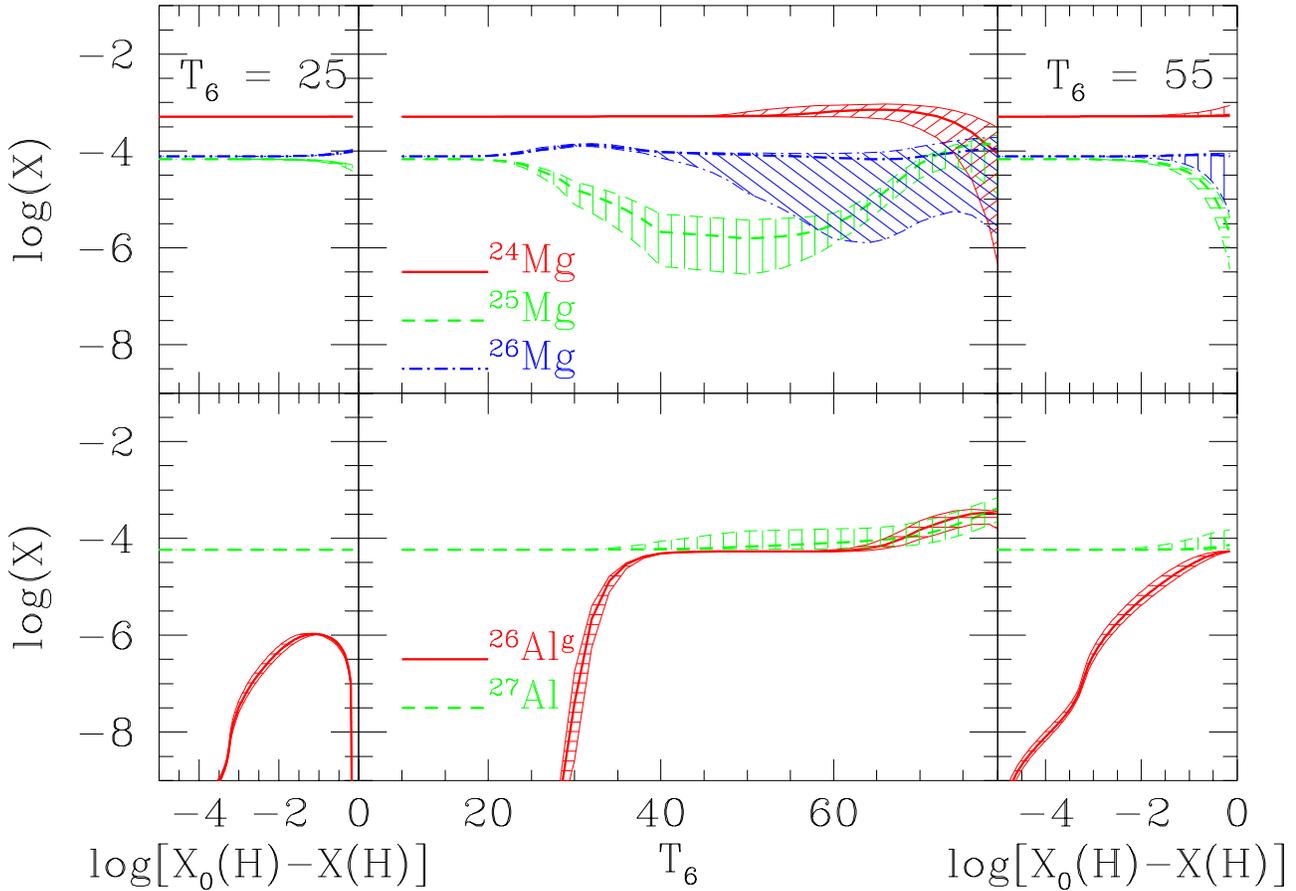}} \hfill
   \caption
        {Same as Fig. \ref{Fig:CNOyields}, but for the nuclides
        involved in the MgAl chain}
   \label{Fig:MgAlyields}
\end{figure*}

It is noticeable that the \chem{Mg}{26} abundance at H exhaustion is
almost temperature independent. This trend differs from the behaviour
of the concentrations of the other Mg and Al isotopes, and results
from two factors. First, the adopted \chem{Mg}{26} proton capture is
slow enough (about ten times slower than prescribed by CF88) for
preventing \chem{Mg}{26} to be destroyed at the considered
temperatures. Second, \chem{Mg}{26} is bypassed by the nuclear flow
associated with the leakage from the NeNa chain at $T_6 \gsimeq
50$. The reaction \reac{Al^g}{26}{p}{\gamma}{Al}{27} is indeed
predicted to be faster than the \chem{Al^g}{26} $\beta$-decay in this
temperature domain.
 
Various aspects of the above analysis may be affected by remaining
rate uncertainties.  In fact, the only proton captures whose rates are
now put on safe grounds are \reac{Mg}{24}{p}{\gamma}{Al}{25} (for
which NACRE and CF88 are in good agreement) and
\reac{Mg}{25}{p}{\gamma}{Al}{26} (for which the NACRE adopted rate is
about 5 times slower than the CF88 one at $T_6 < 80$). In spite of
much recent effort, the other proton capture rates of the MgAl chain
still show more or less large uncertainties in the considered
temperature range, as illustrated in Fig.~\ref{Fig:Alg26pg}.

Due consideration of these uncertainties indicates in particular (see
Fig.~\ref{Fig:MgAlyields}) that, for $T_6 \gsimeq 50$, \chem{Mg}{24}
could be more strongly destroyed than stated above, while
\chem{Mg}{26} could be substantially transformed into \chem{Al}{27} if
the NACRE upper limits on the \reac{Mg}{24}{p}{\gamma}{Al}{25} and
\reac{Mg}{26}{p}{\gamma}{Al}{27} rates were selected. It is also
important to note that the abundances at H exhaustion of
\chem{Al^g}{26} and \chem{Al}{27} are not drastically affected by the
uncertainties left in their proton capture rates, even if these
uncertainties can be quite large (for example, the
\reac{Al^g}{26}{p}{\gamma}{Si}{27} rate is uncertain by more than a
factor of $10^3$ at $T_6 \gsimeq 50$). This situation results from the
fact that even the highest NACRE proton capture rates are not fast
enough for leading to a substantial destruction of the two Al isotopes
by the time H is consumed\footnote{Arnould et al. (1995) have reached
a different conclusion due to a trivial mistake in the
\reac{Al^g}{26}{p}{\gamma}{Si}{27} rate used in their
calculations}. In contrast, the exact conditions under which the MgAl
chain is cycling cannot be reliably specified yet in view of the large
uncertainties still affecting the \reac{Al}{27}{p}{\alpha}{Mg}{24} and
\reac{Al}{27}{p}{\gamma}{Si}{28} rates.

The possibility for the MgAl chain to produce substantial amounts of
\chem{Al^g}{26} is of high interest in view of the prime importance of
this radionuclide in cosmochemistry and $\gamma$-ray line
astronomy. There is now ample observational evidence that
\chem{Al}{26} has been injected live in the forming solar system
before its in situ decay in various meteoritic inclusions (MacPherson
et al.  1995). Its presence in extinct form is also demonstrated in
various types of presolar grains of supposedly circumstellar origin
identified in primitive meteorites (e.g. Zinner 1995). The present-day
galactic plane also contains \chem{Al^g}{26}, as shown by the
observation of a 1.8 MeV $\gamma$-ray line associated with its
$\beta$-decay (e.g.  Prantzos \& Diehl 1996).

The MgAl chain has also a direct bearing on the puzzling Mg-Al
anticorrelation observed in globular cluster red giants. Denissenkov
et al. (1998) have speculated that a strong low-energy resonance could
dominate the rate of \reac{Mg}{24}{p}{\gamma}{Al}{25} at typical cold
H-burning temperatures, and could help explaining these
observations. There is at present no support of any sort to such a
resonant enhancement of this rate.

\section{Helium burning}

The NACRE compilation also provides recommended rates and their lower
and upper limits for most of the $\alpha$-captures involved in the
non-explosive burning of helium. The impact of the remaining rate
uncertainties on the abundances of the elements up to Al affected by
He burning is evaluated in our parametric model for two sets of
conditions: {\it (i)} $\rho = 10^4$~g~cm$^{-3}$ and $T_8=1.5$, adopted
to characterize the central or shell He-burning phases of
intermediate-mass stars ($M\simeq 6$~M$_{\odot}$), and {\it (ii)}
$\rho = 10^4$~g~cm$^{-3}$ and $T_8=3.5$, which can be encountered at
the end of the He burning phase in the core of massive stars or in AGB
thermal pulses. The initial abundances used in these calculations are
adopted as described in Sect.~1.

In contrast to the H-burning case, the abundances during He burning
exhibit some sensitivity to density, as it enters differently the
$3\alpha$ reaction rate and the other $\alpha$-capture
rates. Consequently, the results presented here should not be used to
infer abundances resulting from He burning in specific stellar models,
where the time evolution of the temperature and the density may play
an important role on the final He-burning composition. It has also to
be noted that the neutrons produced by \reac{C}{13}{\alpha}{n}{O}{16}
or \reac{Ne}{22}{\alpha}{n}{Mg}{25} during He burning lead us to
extend the nuclear network to all (about 500) the s-process nuclides
up to Bi.

\begin{figure*}
   \hspace*{3cm}
   \resizebox{13cm}{!}{\includegraphics[angle=-90]{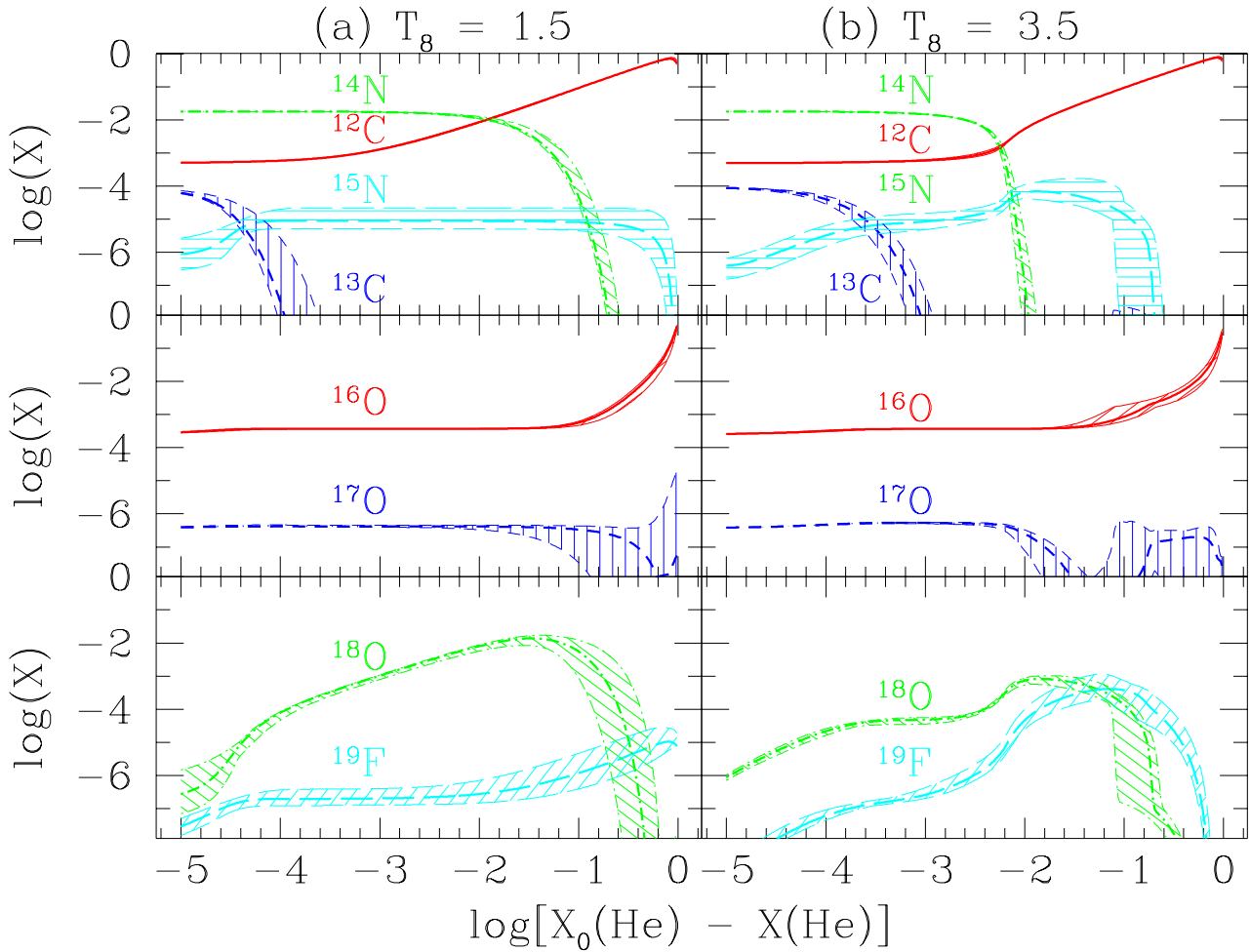}}

   \hfill
   \caption{ 
           Mass fractions of the stable C to F isotopes versus the
           amount of $^4$He burned at constant density
           $\rho=10^4\;{\rm g/cm^3}$ and constant temperature
           $T_8=1.5$ (a: left panel) or $T_8=3.5$ (b: right
           panel). The $^4$He mass fraction is denoted $X$(He), the
           subscript 0 corresponding to its initial value}
   \label{fig_he1}
\end{figure*}

\begin{figure*}
   \hspace*{3cm}
   \resizebox{13cm}{!}{\includegraphics[angle=-90]{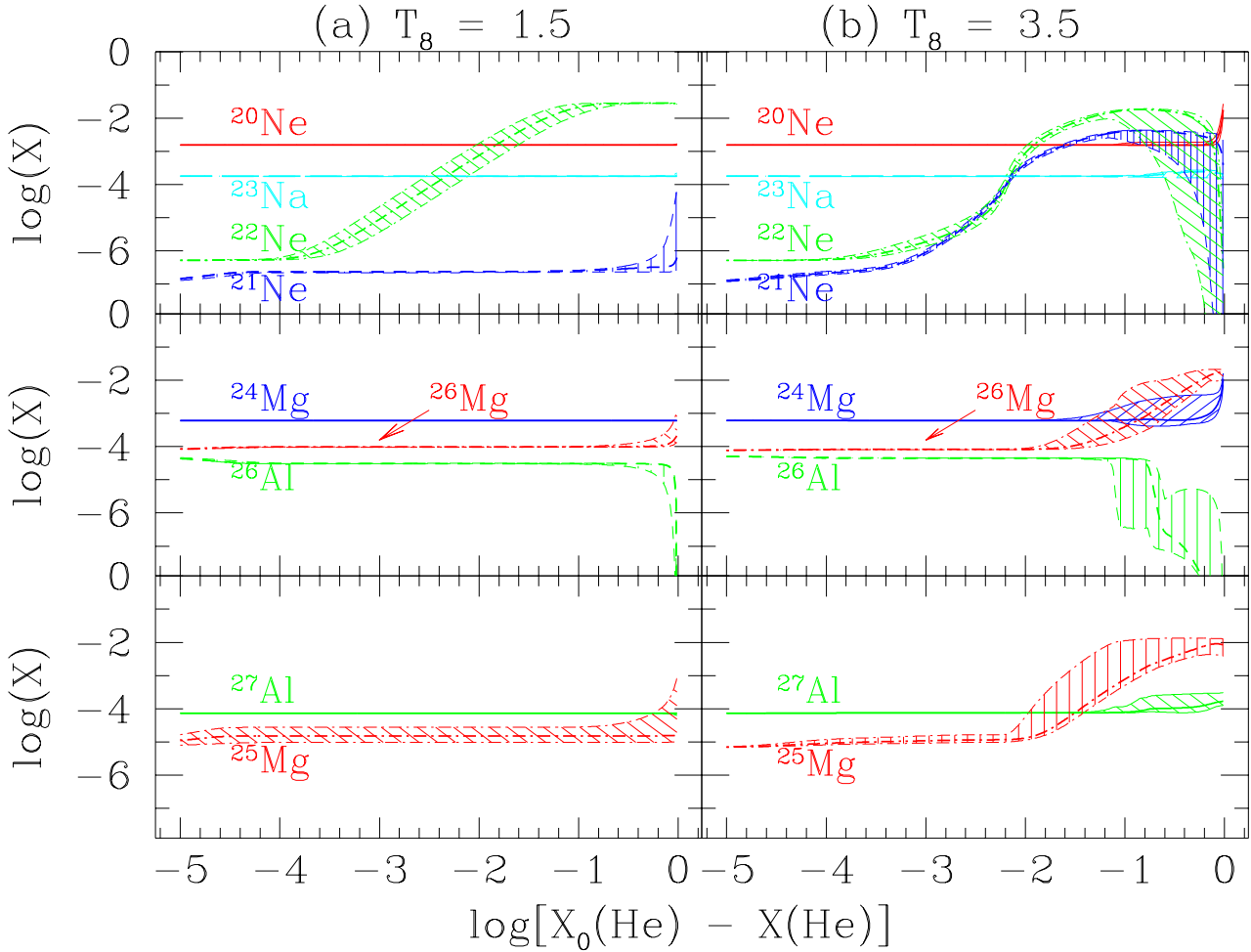}} \hfill
   \caption{Same as Fig.~\ref{fig_he1} for the nuclides from Ne 
     to Al}
   \label{fig_he2}
\end{figure*}

\begin{figure*}
   \hspace*{3cm}
   \resizebox{13cm}{!}{\includegraphics[angle=-90]{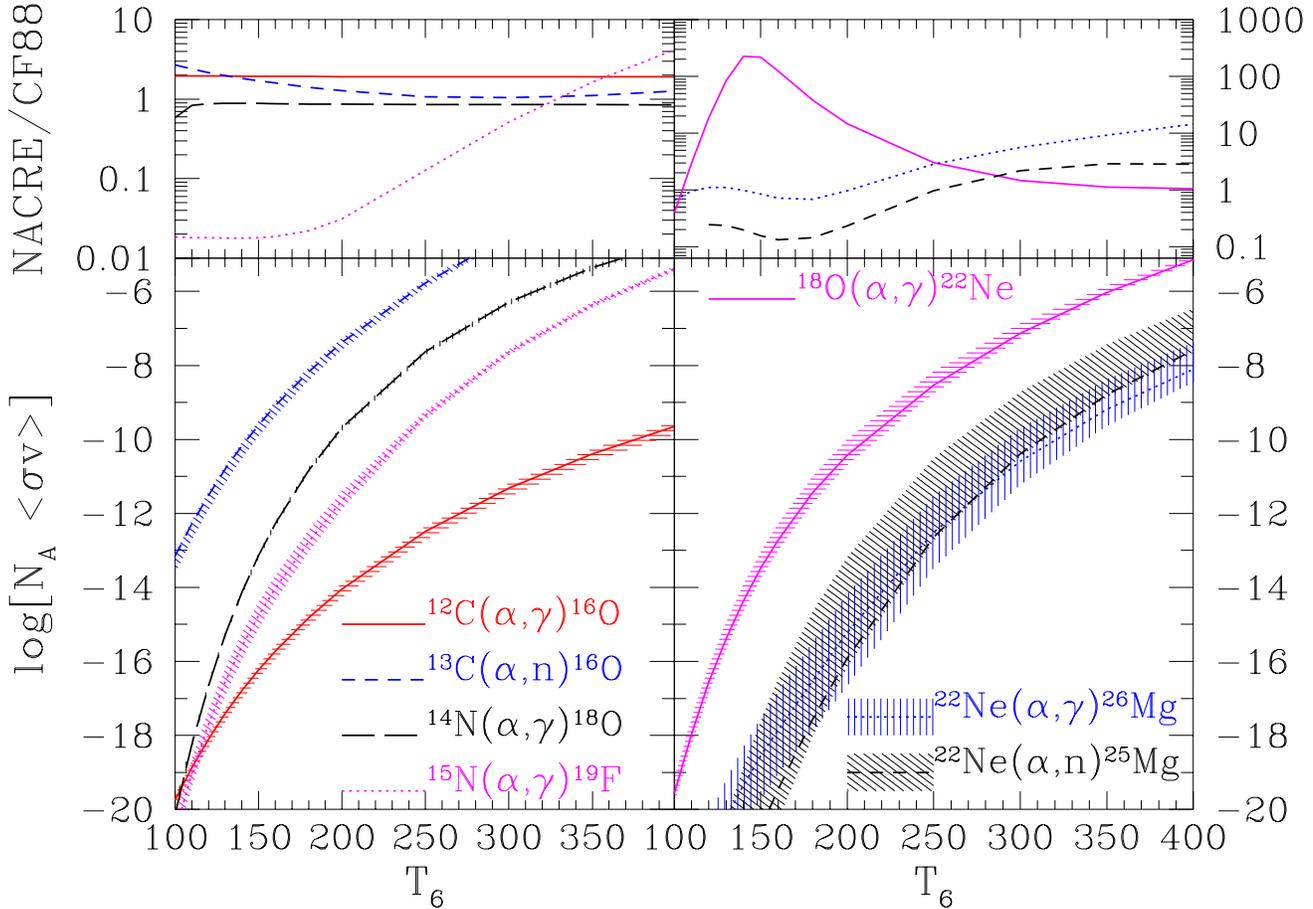}} \hfill
   \caption{Same as Fig.~\ref{Fig:O17pag} for some $\alpha$-capture reactions 
     }
   \label{Fig:rateHe}
\end{figure*}

Figs.~\ref{fig_he1} and \ref{fig_he2} illustrate the evolution during
He burning in the two situations mentioned above of the abundances of
all the stable nuclides between $^{12}{\rm C}$ and $^{27}{\rm Al}$
(plus $^{26}{\rm Al}$).  At low temperature ($T_8 \approx 1.5$;
Figs.~\ref{fig_he1}a and \ref{fig_he2}a), the main reaction flows are

\noindent a) $2\alpha(\alpha,\gamma)^{12}{\rm C}$, followed by
\reac{C}{12}{\alpha}{\gamma}{O}{16} at the very end of He burning. The
factor of 2 uncertainty in the rate of
\reac{C}{12}{\alpha}{\gamma}{O}{16} (Fig.~\ref{Fig:rateHe}) is
responsible for the error bars on the $^{16}{\rm O}$ abundance;

\noindent b) $^{14}{\rm N}(\alpha,\gamma)^{18}{\rm F}(\beta^+)^{18}{\rm O}$, 
followed by 
$^{18}{\rm O}(\alpha,\gamma)^{22}{\rm Ne}$ at the end of He
burning. The resulting $^{22}{\rm Ne}$ does not burn at the considered
low temperature\footnote{In detailed stellar models, the temperature
increases to values in excess of $T_8=3$ towards the end of core (or
shell) He-burning. This may lead to the destruction of $^{22}{\rm Ne}$
by $(\alpha,n)$ (with a concomitant production of neutrons) or
$(\alpha,\gamma)$ reactions, as illustrated on Fig.~\ref{fig_he2}b}.
The uncertainties of a factor of 1.5 and 5 at $T_8=1.5$ in the NACRE
rates of $^{14}{\rm N}(\alpha,\gamma)^{18}{\rm F}$ and $^{18}{\rm
O}(\alpha,\gamma)^{22}{\rm Ne}$, respectively (Fig.~\ref{Fig:rateHe}),
are responsible for the wide range of predicted $^{18}{\rm O}$ and
$^{22}{\rm Ne}$ abundances.  A much larger $^{18}{\rm O}$ abundance at
the end of He burning would result if use were made of the CF88 rate,
which is about 220 times smaller than the NACRE one
(Fig.~\ref{Fig:rateHe}).

\begin{figure}[h]
    \resizebox{\hsize}{!}{\includegraphics[scale=0.30]{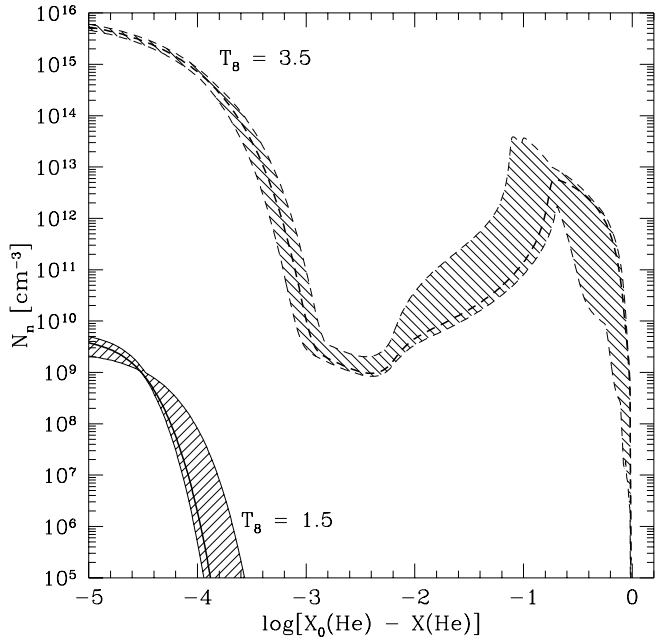}}
   \caption{
         Neutron density versus the amount of helium burned at
         $\rho=10^4\;{\rm g\ cm}^{-3}$ and $T_8=1.5$ (solid line) or
         $T_8=3.5$ (dashed line). The initial \chem{C}{13} mass
         fraction is adopted equal to $10^{-4}$, which is obtained at
         the end of the CNO cycle operating at $T_6 = 60$ and $\rho =
         100$ g cm$^{-3}$ (Sect.~3)}
  \label{fig_he3}
\end{figure}

The neutron density resulting from \reac{C}{13}{\alpha}{n}{O}{16} is
shown in Fig.~\ref{fig_he3}, along with its associated uncertainty.
Albeit small, this neutron irradiation is responsible for the
$^{15}{\rm N}$ and $^{19}{\rm F}$ abundance peaks seen in
Fig.~\ref{fig_he1}a.  They result from $^{14}{\rm
N}(\alpha,\gamma)^{18}{\rm F}(\beta^+)^{18}{\rm O}({\rm
p},\alpha)^{15}{\rm N}(\alpha,\gamma)^{19}{\rm F}$, the protons
originating from $^{14}{\rm N}({\rm n,p})^{14}{\rm C}$. Towards the
end of He burning, $^{19}{\rm F}$ is destroyed by $^{19}{\rm
F}(\alpha,{\rm p})^{22}{\rm Ne}$. Shell He burning in AGB stars or
central He burning in Wolf-Rayet stars have been proposed as a major
site for the galactic production of $^{19}{\rm F}$ (Goriely et
al. 1989; Meynet \& Arnould 1996, 1999; Mowlavi et al. 1998). For AGB
stars, these predictions have been confirmed by the observation of
$^{19}{\rm F}$ overabundances in some of these objects (Jorissen et
al. 1992).  Incomplete He-burning (e.g. in Wolf-Rayet stars) may also
contribute to the galactic enrichment in primary $^{15}{\rm N}$, as
required by the observations of this nuclide in the interstellar
medium (G\"usten \& Ungerechts 1985).

The large \chem{Al}{26} abundance seen on Fig.~\ref{fig_he2}a results
from the particular choice of initial conditions (see Sect.~1), since
\chem{Al}{26} is not produced in the conditions prevailing during
He-burning. Its rapid drop close to the end of He burning results from
the combined effect of $\beta$-decay and $^{26}{\rm Al}({\rm
n,p})^{26}{\rm Mg}$ making use of the few neutrons liberated by
$^{22}{\rm Ne}(\alpha, {\rm n})^{25}{\rm Mg}$.

At higher temperatures (Figs.~\ref{fig_he1}b and \ref{fig_he2}b), the
He-burning nucleosynthesis of the elements up to about Al is
essentially the same as in the low temperature case. The major
differences are observed for \chem{O}{18}, \chem{F}{19},
\chem{Ne}{21}, \chem{Ne}{22}, \chem{Mg}{25}, \chem{Mg}{26} and
\chem{Al}{26}, and are mainly due to a larger neutron production by
\reac{C}{13}{\alpha}{n}{O}{16}, \reac{O}{18}{\alpha}{n}{Ne}{21} and
\reac{Ne}{22}{\alpha}{n}{Mg}{25}. Note that
\reac{O}{18}{\alpha}{n}{Ne}{21} is about 150 times slower than
$^{18}{\rm O}(\alpha,\gamma) ^{22}{\rm Ne}$ in these conditions, but
is fast enough to keep the neutron density above $N_{\rm n}=10^9\;
{\rm cm}^{-3}$ (Fig.~\ref{fig_he3}). These neutrons allow protons to
be produced by the reactions $^{14}{\rm N(n,p)}^{14}{\rm C}$ and
$^{18}{\rm F}({\rm n,p})^{18}{\rm O}$. Additional protons come from
$^{18}{\rm F}(\alpha,{\rm p})^{21}{\rm Ne}$. As a result, $^{15}{\rm
N}$ is produced via $^{18}{\rm F}({\rm n},\alpha)^{15}{\rm N}$,
$^{18}{\rm O}({\rm p},\alpha)^{15}{\rm N}$,\\ $^{14}{\rm N}({\rm
p},\gamma)^{15}{\rm O}(\beta^+)^{15}{\rm N}$ and $^{18}{\rm F}({\rm
p},\alpha)^{15}{\rm O}(\beta^+)^{15}{\rm N}$.  The production of
$^{19}{\rm F}$ follows from $^{15}{\rm N}(\alpha,\gamma)^{19}{\rm
F}$. Since most of the involved reactions have better known rates at
$T_8=3.5$ than at $T_8=1.5$, the corresponding error bars on the
abundances are smaller at higher temperature.  Neutrons are also
responsible for the destruction of any \chem{Al}{26} that may survive
the former H-burning episode.

The operation of $^{22}{\rm Ne}(\alpha,{\rm n})^{25}$Mg at the end of
He burning leads to a non-negligible neutron irradiation which
triggers a weak s-process leading to the overproduction of the $70
\lsimeq A \lsimeq 90$ s-nuclei. Unfortunately, the rate of $^{22}{\rm
Ne}(\alpha,{\rm n})^{25}$Mg remains quite uncertain
(Fig.~\ref{Fig:rateHe}), even at temperatures as high as $T_8=3.5$ (in
this case by a factor of 25). The resulting uncertainty on the neutron
density amounts to a factor of 10 (Fig.~\ref{fig_he3}), while the
total neutron exposure spans the range 0.1 -- 0.3 mbarn$^{-1}$.
Finally, the $\alpha$-captures by the Ne isotopes are fast enough at
temperatures $T_8 > 3$ to alter the Mg isotopic composition. This may
provide a direct observational signature of the operation of the
$^{22}{\rm Ne}(\alpha,{\rm n})^{25}$Mg neutron source in stars
(e.g. Malaney \& Lambert 1988).  Large uncertainties remain, however,
in these reaction rates at He-burning temperatures, except for the
relatively well-determined \reac{Ne}{20}{\alpha}{\gamma}{Mg}{24} rate.

\section{Conclusions}
\label{Sect:Conclusions}

As an aid to the confrontation between spectroscopic observations and
theoretical expectations, the nucleosynthesis associated with the cold
CNO, NeNa and MgAl modes of H burning, as well as with He burning, is
studied with the help of the recent NACRE compilation of nuclear
reaction rates. Special attention is paid to the impact on the derived
abundances of the carefully evaluated uncertainties that still affect
the rates of many reactions. In order to isolate this nuclear effect
in an unambiguous way, a very simple constant temperature and density
model is adopted.

It is shown that large spreads in the abundance predictions for
several nuclides may result not only from a change in temperature, but
also from nuclear physics uncertainties. This additional intricacy has
to be kept in mind when trying to interpret the observations and when
attempting to derive constraints on stellar models from these data.
 
\begin{acknowledgements}
This work has been supported in part by the European Commission under
the Human Capital and Mobility network contract ERBCHRXCT930339 and
the PECO-NIS contract ERBCIPDCT940629.
\end{acknowledgements}

\end{document}